\documentclass[doublecol]{epl2}

\usepackage{graphicx}
\usepackage{amsmath}
\usepackage{amssymb}
\title{Andreev transport in
 two-dimensional normal--superconducting systems in strong magnetic fields}
\shorttitle{Andreev transport in
 two-dimensional normal--superconducting systems \ldots} 

\author{ I.\,M.\,Khaymovich \inst{1}\and N.\,M.\,Chtchelkatchev \inst{2,3,4}\and I.\,A.\,Shereshevskii\inst{1}\and A.\,S.\,Mel'nikov \inst{1}}
\shortauthor{I.M. Khaymovich\etal}
\institute{
  \inst{1} Institute for Physics of Microstructures, Russian Academy of Sciences, 603950 Nizhny Novgorod, GSP-105, Russia\\
  \inst{2} L.D.Landau Institute for Theoretical Physics RAS, 117940 Moscow, Russia\\
  \inst{3} Argonne National Laboratory, Argonne, IL 60439, USA\\
  \inst{4} Institute for High Pressure Physics, Russian Academy of Sciences, Troitsk 142190, Moscow region, Russia
  }
  {Theory of electronic transport; scattering
mechanisms}   {Electronic transport in mesoscopic
systems}   {Electronic transport in nanoscale
materials and structures}   {Tunneling phenomena;
point contacts, weak links, Josephson effects}

\abstract{The conductance in two--dimensional (2D)
normal--superconducting (NS) systems is analyzed in the limit of
 strong magnetic fields when the  transport is
mediated by the electron--hole states bound to the sample edges
and NS interface, i.e., in the Integer Quantum Hall Effect regime.
 The Andreev--type process of the conversion of the quasiparticle
current into the superflow is shown to be strongly affected by the
 mixing of the edge states localized at the
NS and insulating boundaries.  The magnetoconductance in 2D NS
structures is calculated for both quadratic and Dirac--like normal
state spectra. Assuming  a random scattering of the edge modes we
analyze both the average value and fluctuations of conductance for
an arbitrary number of conducting channels. }

\begin{document}



\maketitle


\section{Introduction}
Andreev transport phenomena, i.e.,  transport effects associated
with the conversion of electrons into holes, are known to
determine the distinctive features of a wide class of hybrid
structures consisting of the normal (N) and superconducting (S)
metal parts (see, e.g., \cite{been_rand_mat} and references
therein). Applying rather high magnetic fields one can drastically
affect the physics of these Andreev--type effects due to a strong
modification of the transport mode structure
\cite{Shon,Chtch-1,AkhmerBeen_valley_polar,Xie_Sun} which is
typical for the systems in the Integer Quantum Hall Effect (IQHE)
regime. Provided the radii of the cyclotron orbits in the normal
part of the system become less than the mean free path the
transport appears to be determined by the waves bound to the
sample edges. Depending on the momenta of these waves and
quasiparticle charge the transport modes are localized near the
different edges and, thus, the wavefunctions of the incoming and
outgoing particles appear to be spatially separated. Thus, the
magnetic field destroys the basic backscattering property of the
standard Andreev reflection.

\begin{figure}[b]
\center\includegraphics[width=0.46\textwidth]{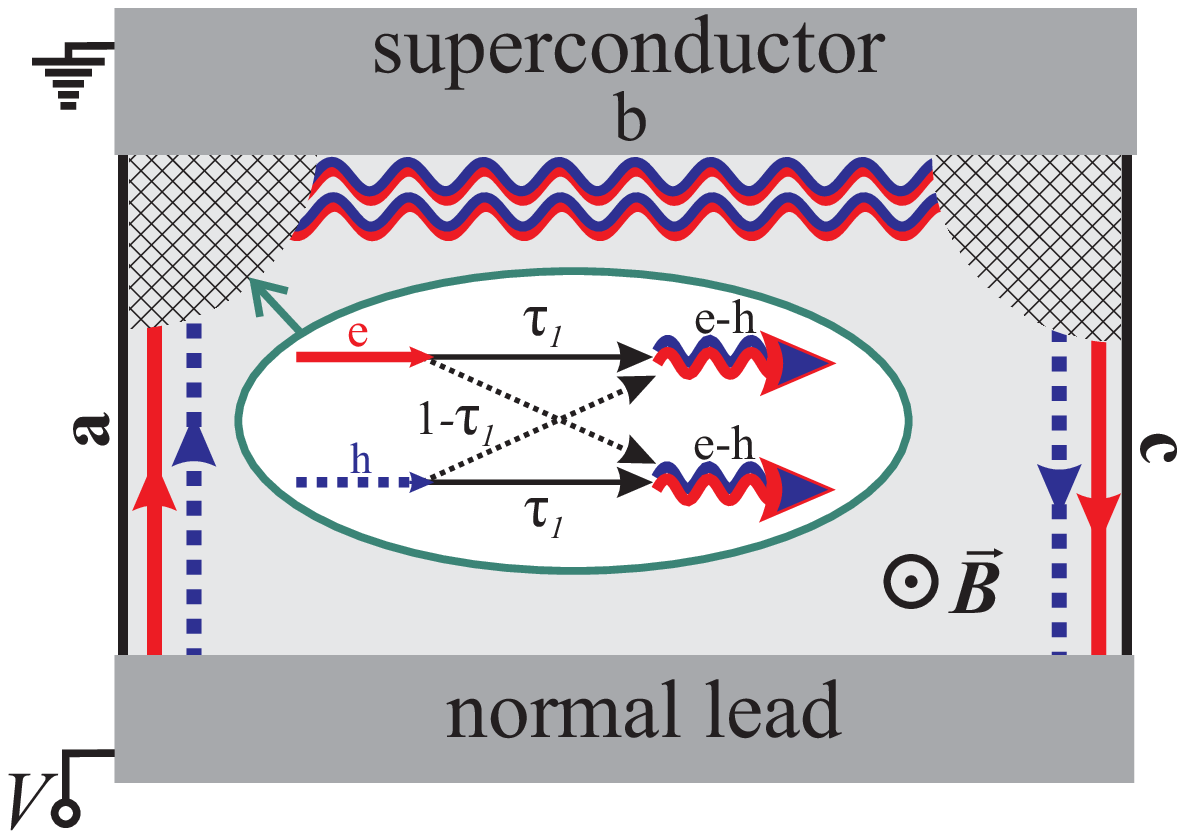}
\caption{\label{setup} (Color online) The junction between the
superconductor  and the chiral metal  [edge states in high
mobility 2D metal in the IQHE regime]. Solid (dotted) lines
correspond to the electron (hole) edge states at ``a'' and ``c''
insulating boundaries,  and wavy lines correspond to the
hybridized electron-hole modes at the superconducting boundary
``b''. Inset: Edge mode scattering from insulating edge ``a'' to
superconducting edge ``b'' with probabilities $\tau_1$ and
$(1-\tau_1)$.}
\end{figure}

 It is the goal of the present work to suggest a general
theoretical description of the Andreev  transport mediated by the
edge states in the IQHE regime. We consider here an exemplary
two--dimensional (2D) NS  system shown in Fig.~\ref{setup}. Such
type of mesoscopic junctions based on a 2D electronic gas (2DEG)
or gapless 2D semiconductors like graphene are in the focus of
current experimental and theoretical research
\cite{Shon,Chtch-1,AkhmerBeen_valley_polar,Xie_Sun,many-1,many-2,many-3,many-4,many-5}.
To elucidate our main results we start here from a qualitative
description of the transport mediated by the edge states.  An
electron injected from the normal conductor goes to the
superconductor through an edge state ``a''. At the ``ab''-corner
it transforms into two types of hybridized electron-hole states at
the boundary "b" with the probabilities $\tau_1$ and $(1-\tau_1)$,
respectively (see the inset of Fig.~\ref{setup}). Similarly to the
situation at the ``ab''-corner each of electron-hole
quasiparticles transforms with the probabilities $\tau_2$ and
$(1-\tau_2)$ into an electron and a hole which return to the
normal lead through the edge states "c". Without mode mixing
($\tau_{1,2} = 0$ or $1$) each initial electron (hole) state at
the boundary "a" completely transforms into the final electron
(hole) state at the boundary "c" and the probability of the
electron--hole conversion is zero, so that the total conductance
$G$ vanishes.
Taking account of the quasiparticle mode mixing at the corners we
find:
\begin{equation}\label{Both:G_aver}
G = 2G_0 N\left[\tau_1(1 - \tau_2) + \tau_2(1 - \tau_1)\right] \ ,
\end{equation}
where $G_0 = e^2/\pi\hbar$ is the conductance quantum, and $N$ is
the number of propagating electron edge states. Thus, it is the
mode mixing which is responsible for the conversion of the
quasiparticle current into the supercurrent outgoing from the NS
boundary.

Generally, the solution of the problem of coupling between the
edge states near the corner
  is rather
complicated and depends on the details of the system geometry.
Therefore only limiting cases have been previously considered: (i)
quasiclassical limit with the large number of the edge states,
when the problem can be treated considering the particles and
holes on the cyclotron orbits skipping along the surface
\cite{Chtch-1}; (ii) quantum limit when the number of the edge
states is of the order of unity
\cite{AkhmerBeen_valley_polar,Xie_Sun}. The first case corresponds
to rather large Fermi energies comparing to the Landau level
spacing. The corresponding oscillating behavior of the conductance
of the 2D NS junction vs magnetic field, junction width and/or
Fermi level has been analyzed in detail in
Refs.~\cite{Chtch-1,Shon}.

In this Letter we analyze the magnetoconductance behavior in the
NS structures for an arbitrary number of transport modes taking
into account the mixing of the edge states near the corners.
Adopting a phenomenological description of the mode mixing problem
based on the transfer matrix approach we find a simple expression
for the conductance in the quantum limit. Such model suggests a
simple explanation of the oscillatory phenomena mentioned above
and brings out their dependence on the mode coupling parameters.
To find the conductance for an arbitrary number of quantum
channels we assume the mode mixing at the corners to be random in
the sense that the appropriate scattering matrices are uniformly
distributed (see below for details).  Such approach allows us to
find universal expressions for the average conductance and its
fluctuations which depend only on fundamental constants and number
of transport channels:
\begin{equation}
\label{Both:G_universal}
    \langle G\rangle=G_0 N \ ,
\end{equation}
\begin{equation}\label{Both:sigma_universal}
    \sigma_G = \sqrt{\left<G^2\right>-\left<G\right>^2}=
G_0\sqrt{\frac{N}{4(2N+1)}} \ .
\end{equation}

In the equation~\eqref{Both:G_universal} contribution of each
electron transport mode to the conductance equals to the
conductance quantum $G_0$ instead of $2G_0$ in ballistic
NS-junctions without magnetic field. Such conductance reduction is
caused by the levelling of outgoing (along the ``c''-edge)
electron and hole probabilities in the limit of strong disorder,
which saturate at the value 1/2. Such levelling is analogous to
the one observed in numerical simulations in \cite{Xie_Sun} for
the tight-binding model with the disorder in on-site energies.

\section{Basic equations and edge state spectra} The
spectra of quasiparticle edge states can be found using the
Bogolubov--de Gennes (BdG) equations written for electron--like
($u$) and hole--like ($v$) parts of the wave function $\hat\Psi =
(\hat u, \hat v)$:
\begin{equation}\label{BdG-equation}
\left(
\begin{array}{cc}
\hat H - \mu & \Delta\\
\Delta^* & \mu - \mathcal{T}\hat H \mathcal{T}^{-1}
\end{array}
\right)\hat\Psi = E \hat\Psi \ .
\end{equation}
Here $\mathcal{T}$ is the time--reversal operator, $\Delta$ is the
gap operator, and the energy $E$ is measured relative to the Fermi
level $\mu$. Note that we neglect here the Zeeman shift of the
quasiparticle spectra caused by the interaction of magnetic field
with the true electron spin.


For 2DEG the single particle hamiltonian $\hat{H}$ takes the
Schr\"{o}dinger form $\hat H({\bf A}) = ({\bf p}-\frac{e}{c}{\bf
A})^2/2m $, where ${\bf p} = -i\hbar({\partial}/{\partial
x};\partial/\partial y)$ is the momentum operator in the x-y plane
of the 2D system, and ${\bf A}$ is the vector potential
corresponding to the  magnetic field ${\bf B}$ perpendicular to
the system plane. In graphene there appear 2 sublattice
(pseudospin) and 2 valley (isospin) degrees of freedom and in the
"valley-isotropic" basis the hamiltonian  could be written as
follows: $\hat{H} = v_F\tau_0\otimes\sigma\left({\bf
p}-\frac{e}{c}{\bf A}\right)$ (see, e.g.,
\cite{AkhmerBeen_valley_polar}). Here
  $v_F$ is the Fermi velocity, $\sigma_i$ and
$\tau_i$ are  the Pauli matrices  acting in the sublattice and
valley spaces, respectively, and  $\sigma_0$, $\tau_0$ are the
$2\times2$ unit matrices.

The boundary conditions and corresponding edge state spectra at
the boundaries with isolator and superconductor have been
previously studied for 2DEG \cite{Shon} as well as for graphene
\cite{AkhmerBeen_valley_polar,volkov,levitov,burset}. The boundary
condition at the NS interface with 2DEG  couples electron $u$ and
hole $v$ parts of wave function and in quasiclassical limit takes
the usual form $v = e^{- i\beta}u$, where $\beta =\arccos
(\varepsilon/\Delta)$. The wave function at the 2DEG insulating
edge vanishes $\Psi = 0$.
 According to Akhmerov and
Beenakker  \cite{AkhmerBeen_valley_polar} the graphene-isolator
boundary conditions
\begin{equation}\label{Graphene:BC_GI}
\Psi = \left({\bf\nu},{\bf \hat \tau}\right)\otimes\left({\bf
n_{\bot}},{\bf \hat\sigma}\right)\Psi
\end{equation}
are crucially determined by the  isospin vector ${\bf \nu}$ while
the resulting quasiparticle spectrum depends on the vector $\bf
n_{\bot}$. Here the unit vector $\bf n_{\bot}$ should have a zero
projection on the direction normal to the graphene edge. The
graphene-superconductor (GS) interface boundary condition  doesn't
depend on valley degree of freedom and for subgap energies it
could be written as follows: $\hat v = e^{- i\beta({\bf n\cdot
\sigma})}\hat u$. Thus, the edge state spectrum is valley
degenerate.

Taking the case of a homogeneous boundary and choosing the gauge
with the vector potential parallel to the boundary one can find a
set of spectral branches $E_n(k_\parallel)$ vs the conserved
momentum component $k_\parallel$ along the interface. Here we
introduce an integer index $n$ enumerating the branches. Thus,
each insulating edge (``a'' or ``c'') supports $2N$ propagating
edge modes: $N$ electron-like modes and $N$ hole-like modes. The
NS interface also supports $2N$ propagating (valley degenerated in
graphene case) modes with mixed electron-hole wave functions.

\section{Mixing of the edge modes. Transfer matrix
approach} Considering the transport mediated by the edge states we
use a simple phenomenological model based on the transfer matrix
approach. We introduce a transfer matrix $\hat S$ which couples
the quasiparticle edge waves propagating along the ``a'' and ``c''
boundaries. This matrix calculated for the states at the Fermi
level is known to determine the linear transport characteristics
at zero temperature \cite{been_rand_mat,Blonder}. It is important
to note here that at the Fermi level the matrix $\hat S$ appears
to be simultaneously a scattering matrix coupling the incoming and
outgoing electron--hole waves. Indeed, in this case all the states
propagating along the interfaces in Fig.~\ref{setup} have the same
sign of the group velocity $\partial E_n/\partial k_\parallel$
\cite{Shon,AkhmerBeen_valley_polar} and, thus, all quasiparticle
fluxes are flowing clockwise for a chosen magnetic field
direction.

The BdG equations (\ref{BdG-equation}) are known to be invariant
with respect to the transformation converting electrons into holes
and changing the sign of energy $E$ and, as a consequence, all the
edge modes can be divided into two groups connected by this
transformation. For the NS interfaces we denote these groups as
$g_+$ and $g_-$ while for the boundaries with an insulator (``a''
and ``c'') these groups just coincide with pure electron ($u_a$
and $u_c$) and hole ($v_a$ and $v_c$) waves. Each of hybrid states
$g_{\pm}$  propagating along the ballistic NS boundary ``b'' of
the length $L$ acquires a phase factor $e^{\pm i k_n L}$, where
$k_n$ is the momentum satisfying the equation $E_n(k_n)=0$ for the
spectral branches at the NS boundary. The scattering processes at
the corners  ``ab'' and ``bc'' (see Fig.~\ref{setup}) could be
described by unitary transfer matrices $\hat{T}_{1,2}$, which
couple the incident and transmitted quasiparticle waves:
\begin{equation}
\left(g_+\atop g_-\right)=\hat{T}_{1}\left(u_a\atop v_a\right) \ ,
\quad \left(u_c\atop v_c\right)=\hat{T}_{2} \left(g_+\atop
g_-\right)\ ,
\end{equation}
where $u_a$, $v_a$, $u_c$, $v_c$, $g_+$, $g_-$ are the sets of
wave amplitudes corresponding to the solutions of BdG equations at
the Fermi level. The unitarity of the matrices $\hat T_{1,2}$ is a
consequence of the quasiparticle current conservation. The
scattering matrices $\hat{T}_{1,2}$ can be conveniently presented
in the four--block form:
\begin{equation}\label{Both:Scat_mat_T_1}
\hat T_1 =\left(\begin{array}{cc}
\hat t_{+e}& \hat t_{+h}\\
\hat t_{-e}& \hat t_{-h}
\end{array}
\right) \ , \quad
\hat T_2 =\left(\begin{array}{cc}
\hat t_{e+}& \hat t_{e-}\\
\hat t_{h+}& \hat t_{h-}
\end{array}
\right) \ .
\end{equation}
The total transfer matrix $\hat S$ can be written as a product of
three matrices describing subsequent scattering and propagation
processes discussed above:
\begin{equation}\label{Both:Scat_mat_S}
\hat S =\left(\begin{array}{cc}
\hat s_{ee}& \hat s_{eh}\\
\hat s_{he}& \hat s_{hh}
\end{array}
\right) = \hat T_2 \hat\Lambda_L \hat T_1 \ ,
\end{equation}
where $\hat{\Lambda} = {\rm diag}\left(e^{i k_1 L},\ldots,e^{i
k_{N} L},e^{-i k_1 L},\ldots,e^{-i k_{N} L}\right)$ is a diagonal
transfer matrix of phase factors acquired at the NS boundary. The
blocks $\hat s_{ee},\hat s_{eh},\hat s_{he},\hat s_{hh}$ of the
matrix $\hat{S}$ describe the scattering between the edge states
at the ``a'' and ``c'' boundaries. The zero--temperature
conductance
is given by following expression \cite{Blonder}:
\begin{equation}\label{Both:G_TE}
G = G_0{\rm
Sp}\left(\hat{I}-\hat{s}_{ee}^+\hat{s}_{ee}+\hat{s}_{eh}^+\hat{s}_{eh}\right)
= 2G_0{\rm Sp}\left(\hat{s}_{eh}^+\hat{s}_{eh}\right) \ .
\end{equation}

\section{Magnetoconductance in NS junctions. Quantum
limit} At sufficiently small Fermi energy $\mu$ each edge supports
only two propagating modes
and each block of scattering matrices becomes a single complex
amplitude. It is convenient to parametrize these amplitudes as
follows
\begin{equation}
\label{2DEG:t_+e}t_{+e} =  \sqrt{1-\tau_1}e^{i (\varphi_1 +
\theta_1/2)} \ ,\,
 t_{-e} = \sqrt{\tau_1}e^{i (\varphi_1 -
\theta_1/2)}\ ,
\end{equation}
\begin{equation}
\label{2DEG:t_e+} t_{e+}  = \sqrt{1-\tau_2}e^{i (\varphi_2 +
\theta_2/2)}\ ,\, t_{e-} = \sqrt{\tau_2}e^{i (\varphi_2 -
\theta_2/2)}\ ,
\end{equation}
$t_{\mp h} =\pm t_{\pm e}^*$, $t_{ h\mp} =\pm t_{ e\pm}^*$,  where
$\tau_1$ is the probability that electron at the boundary ``a''
scatters into the second type of hybrid modes, $\tau_2$ is the
probability that the first type hybrid mode scatters into the hole
at the boundary ``c''. The matrix of phase factors takes the form:
$\hat{\Lambda} = {\rm diag}\left(e^{i k_1 L},e^{-i k_1 L}\right)$,
where $k_1$ is the momentum value at which the spectral branch at
the NS boundary crosses the Fermi level. Omitting the calculation
details we present here the final expression for the 2DEG --
superconductor junction conductance:
\begin{equation}\label{2DEG:G_TE}
G = 2G_0(p_{21}+p_{12}-2\sqrt{p_{12}p_{21}}\cos\gamma) \ ,
\end{equation}
where $p_{nm}=\tau_n(1-\tau_m)$, $\gamma = 2k_1
L+\theta_1+\theta_2$. In the symmetric geometry of
Fig.~\ref{setup}, i.e., when the ``ab'' and ``bc'' corners have
the same shape, one can expect the appearance of an additional
symmetry of scattering matrices describing the mode mixing:
$\hat{T}_2 = \hat T_1^T$ ($p_{12}=p_{21}$). The expression for
conductance in this case can be further simplified:
\begin{equation}\label{2DEG:G_TE_T2=T1^T}
G = 4G_0\tau(1-\tau)\left(1-\cos\gamma\right) \ ,
\end{equation}
where $\tau_1 = \tau_2 \equiv \tau$. The conductance reveals an
oscillating behavior vs the junction width $L$ which is, in fact,
a consequence of quantum mechanical interference of the edge waves
propagating along the NS boundary. The expression
(\ref{2DEG:G_TE_T2=T1^T}) is in good agreement with the
qualitative arguments in the introduction: the absence of mode
mixing corresponding to the limits $\tau=0$ or $\tau=1$ causes a
complete suppression of the charge transport.

Considering a quantum limit for GS junctions we need to emphasize
two important distinctive features. First, the momentum of the
zero energy mode ($E_1(k_1)=0$) at the GS boundary appears to
vanish ($k_1=0$) and both states ($g_+$ and $g_-$) become
degenerate. Second, the scattering matrices $\hat{T}_{1,2}$
crucially depend on the isospin degree of freedom. Following
Ref.\cite{AkhmerBeen_valley_polar} we introduce the isospin
operator eigenvectors' basis $\left({\bf \nu}_m,{\bf
\tau}\right)\left|\pm\nu_m\right> = \pm\left|\pm\nu_m\right>$,
where $\nu_{1,2}$ are the isospin vectors characterizing the ``a''
and ``c'' boundaries (see the boundary condition
(\ref{Graphene:BC_GI})). The scattering matrices take the form:
\begin{multline}
\hat T_1 = \left|\nu_1\right>\left<\nu_1\right|t_{+e} +
\left|-\nu_1\right>\left<\nu_1\right|t_{-e} +\\
+\left|\nu_1\right>\left<-\nu_1\right|t_{+h} +
\left|-\nu_1\right>\left<-\nu_1\right|t_{-h}
\end{multline}
\begin{multline}
\hat T_2 = \left|\nu_2\right>\left<\nu_2\right|t_{e+} +
\left|-\nu_2\right>\left<\nu_2\right|t_{h+} +\\
+\left|\nu_2\right>\left<-\nu_2\right|t_{e-} +
\left|-\nu_2\right>\left<-\nu_2\right|t_{h-}
\end{multline}
Introducing the notation $\chi=\widehat{({\bf \nu}_1,{\bf
\nu}_2)}$ for
 the angle between the isospin
vectors one can get the conductance of the system in the form:
\begin{multline}\label{Graphene:G_TE}
G =
2G_0\left[\sin^2\tfrac{\chi}{2}+\left(p_{21}+p_{12}\right)\cos\chi
\right.\\
-2\sqrt{p_{12}p_{21}}(\cos\theta_1\cos\theta_2-\sin\theta_1\sin\theta_2\cos\chi)\\
-\sin\chi\left((1-2\tau_1)\sqrt{p_{22}}\sin\theta_1\left.+
(1-2\tau_2)\sqrt{p_{11}}\sin\theta_2\right)\right] \ .
\end{multline}
 Assuming a symmetric geometry of Fig.~\ref{setup} we put
$\hat{T}_2 = \hat T_1^T$ and find:
\begin{multline}\label{Graphene:G_TE_T2=T1^t}
G = 2G_0\left(2\sqrt{p}\cdot\cos\tfrac{\chi}{2}\sin\theta-
(1-2\tau)\sin\tfrac{\chi}{2}\right)^2 \ ,
\end{multline}
where $\theta_1 = \theta_2 \equiv \theta$, $p = \tau(1-\tau)$.
Contrary to the 2DEG case the conductance does not depend on the
junction width $L$ which is a natural consequence of zero phase
acquired by the waves propagating along the GS boundary in the
two--mode limit. Another new feature specific for the case of
graphene is that the additional mode mixing occurs due to the
mismatch of the isospin directions at different insulating
boundaries. Note that neglecting the intervalley scattering at the
corners we put $\tau=0$ or $\tau=1$ and get the limit considered
in Ref. \cite{AkhmerBeen_valley_polar}.

\section{Magnetoconductance in NS junctions.
Random--matrix theory} Considering the charge transport mediated
by a large number of edge states it is natural to expect that the
conductance will be given by the sum of phase factors:
\begin{equation}
G=\sum_{ij} a_{ij} e^{i(k_i-k_j)L} \ ,
\end{equation}
where the hermitian matrix $a_{ij}$ is determined by the transfer
matrix parameters. The  oscillating behavior of conductance vs $L$
becomes more complicated than in a two--mode limit and is
generally characterized by a set of incommensurable periods.
Previously these oscillations have been predicted on the basis of
quasiclassical method for quasiparticles moving along skipping
cyclotron orbits \cite{Chtch-1}. In real experimental situation
the mode interference and the corresponding oscillations should
be, of course, smeared due to the effect of sample imperfections,
e.g., roughness, etc. Here we suggest a phenomenological approach
to treat the problem taking account of these effects for an
arbitrary number of modes. We model the scattering caused by the
sample imperfections introducing random transfer matrices $\hat
T_{1,2}$ and applying standard methods for calculation of the
ensemble averages \cite{been_rand_mat}.

The symmetry properties of  the transfer matrix $\hat T_m$ allow
us to introduce a polar decomposition (cf. \cite{Mello,Martin}): $
\hat{T}_{m} = \hat \alpha_m \hat \Theta_m \hat \beta_m$, where
$$\hat \alpha_m = \left(\begin{array}{cc}
\hat A_m & 0\\
0 & \hat A_m^*
\end{array}\right),  \, \hat\beta_m = \left(\begin{array}{cc}
\hat{B}_m & 0\\
0 & \hat{B}_m^*
\end{array}\right) \ ,$$
$$\hat \Theta_m = \left(\begin{array}{cc}
\sqrt{1-\hat{\tau}_m} & \sqrt{\hat{\tau}_m}\\
-\sqrt{\hat{\tau}_m} & \sqrt{1-\hat{\tau}_m}
\end{array}\right) \ ,$$
and the unitary matrices $\hat A_m, \hat B_m$ characterize the
scattering phase shifts while the
 diagonal matrix $\hat\tau_m = {\rm
diag}(\tau_m^{(1)},\ldots,\tau_m^{(N)})$ consists of the
eigenvalues $\tau_{1}^{(n)}$ ($\tau_{2}^{(n)}$) of the matrix
$\hat t_{-e}^+\hat t_{-e}$ ($\hat t_{e-}^+\hat t_{e-}$). The
values $\tau_{m}^{(n)}$ give us the probabilities of transitions
between the modes. Any scattering $2N\times 2N$-matrix $\hat X$
invariant under electron-hole converting transformation of BdG
equations belongs to the compact symplectic group ${\rm Sp}(N) =
\{\hat X _{ 2N\times 2N}:\hat X\hat X^+=\hat I, \hat J\hat X+\hat
X^T\hat J=0\}$, where
$$\hat J = \left(\begin{array}{cc}0&\hat I\\-\hat I&0\end{array}\right)$$
Our further calculations are based on the simplest assumption
about the distributions of the random transfer matrices: we
consider the random unitary matrices $\hat T_m$ uniformly
distributed in the compact symplectic group ${\rm Sp}(N)$.

The uniform distribution of an element $\hat A$ of  the compact
group is defined with respect to a measure $d\mu(\hat A)$ which is
invariant under multiplication: $d\mu(\hat A) = d\mu(\hat U\hat
A\hat V)$ for arbitrary elements $\hat U,\hat V$ belonging to this
group. This measure is known as the "invariant measure" or "Haar
measure" \cite{been_rand_mat}. Note that under such assumption all
distinctive features of the graphene case associated with the
isospin degree of freedom  do not reveal in the averages and,
thus, further expressions are valid for both types of junctions
under consideration.

Analogously to Ref.~\cite{Baranger} we derive a full distribution
of $\hat T_m$ for arbitrary $N$ and the statistics of the
eigenvalues $\{\tau_m\}$
\begin{equation}
d\mu(\hat T_m) = P(\{\tau_m\})\prod\limits_a d\tau_m^{(a)}
d\mu(A_m)d\mu(B_m) \ ,
\end{equation}
where $P(\{\tau_m\})=C\prod\limits_{i\neq
j}|\tau_m^{(i)}-\tau_m^{(j)}|$ is the joint probability
distribution of the $\{\tau_m\}$ values,
 $d\mu(A_m)$ [$d\mu(B_m)$] is the
invariant (Haar's) measure on the unitary group for matrix $\hat
A_m$ [$\hat B_m$] and $C$ is a normalization constant.

The conductance averaged over the matrices $\hat A_m$, $\hat B_m$
takes the form:
\begin{equation}
\left<G\right>_{ A, B} = 2G_0 N\left[\tilde\tau_1(1 -
\tilde\tau_2) + \tilde\tau_2(1 - \tilde\tau_1)\right] \ ,
\end{equation}
where $\tilde\tau_p = {\rm Sp}(\hat\tau_p)/N$. This expression
gives us a generalization of the Eq.~\eqref{2DEG:G_TE} averaged
over the phase $\gamma$ and written for an arbitrary number $2N$
of transport modes. After averaging over the
 eigenvalues $\tau_m^{(n)}$ we find the expressions
 \eqref{Both:G_universal} and \eqref{Both:sigma_universal}
 for the conductance and  its square deviation.
The ensemble average conductance is proportional to the number of
channels and doesn't depend on junction width $L$. The square
deviation increases with the growing channel number and saturates
at the universal number $G_0/2\sqrt{2}$.

Changing the applied magnetic field $\bf B$ one could control the
number of edge modes which decreases stepwise with the increasing
field.

Thus, the dependence of conductance vs the inverse magnetic field
$B^{-1}$ reveals a series of equidistant steps
$$G_{2d} = G_0\left\lfloor\frac{m c \mu}{e \hbar B}-\frac{1}{2}\right\rfloor$$
$$G_{gr} = G_0\left\lfloor\frac{c \mu^2}{2 v_F^2 e \hbar B}\right\rfloor$$
for 2DEG and graphene respectively. Here we introduce the notation
$\lfloor...\rfloor$ for the integer part.


\acknowledgements We are thankful to M.\ A.\ Silaev for many
stimulating discussions. This work was supported in part by the
Russian Foundation for Basic Research, the "Dynasty" Foundation,
Russian Presidential Program Grant No. MK-7674.2010.2, and the FTP
"Scientific and educational personnel of innovative Russia in
2009-2013".

\end{document}